\documentclass[12pt, article,oneside, 
headinclude,
footinclude, 
BCOR=5mm]{scrbook}
\usepackage{setspace} 
 \makeatletter
\DeclareOldFontCommand{\rm}{\normalfont\rmfamily}{\mathrm}
 \DeclareOldFontCommand{\sf}{\normalfont\sffamily}{\mathsf}
 \DeclareOldFontCommand{\tt}{\normalfont\ttfamily}{\mathtt}
 \DeclareOldFontCommand{\bf}{\normalfont\bfseries}{\mathbf}
 \DeclareOldFontCommand{\it}{\normalfont\itshape}{\mathit}
 \DeclareOldFontCommand{\sl}{\normalfont\slshape}{\@nomath\sl}
 \DeclareOldFontCommand{\sc}{\normalfont\scshape}{\@nomath\sc}
 \makeatother
\usepackage[english]{babel}
\usepackage[T1]{fontenc} 
\usepackage{times}
\usepackage[utf8]{inputenc}
\usepackage{newunicodechar}
\newunicodechar{〈}{\langle}
\usepackage{tabularx}
\usepackage{graphicx} 
\graphicspath{{Figures/}} 

\usepackage{enumitem} 
\newlist{steps}{enumerate}{1}
\setlist[steps, 1]{label = Step \arabic*:}

\usepackage{lipsum} 

\usepackage{amsmath,amssymb,amsthm} 
\usepackage{varioref} 
\usepackage{booktabs} 

\usepackage{siunitx} 

\usepackage{pgfplotstable} 

\usepackage{etoolbox}

\usepackage{float}

\usepackage{amsmath}

\usepackage[T1]{fontenc}

\usepackage{natbib}

\usepackage{listings}

\usepackage[labelfont=bf]{caption}
\usepackage{longtable}

\usepackage{filecontents}  

\usepackage{amsfonts}

\usepackage[colorinlistoftodos]{todonotes}

\usepackage{algorithm,algpseudocode}
\usepackage{algorithmicx}

\usepackage{lineno}

\usepackage{etoolbox}

\usepackage{blindtext}

\usepackage{enumitem}
\usepackage{lmodern}
\usepackage{fullpage}
\usepackage{alltt}
\usepackage{boxedminipage}
\usepackage{color}

\usepackage{booktabs}

  \usepackage{url}
\usepackage{subfig}
\usepackage[font={normal,it}]{caption}

\theoremstyle{definition} 

\theoremstyle{plain} 

\theoremstyle{remark} 
\usepackage{authblk}
\usepackage{chngcntr}
\counterwithin{figure}{section}
\PassOptionsToPackage{hyphens}{url}
\usepackage{breakcites}
\usepackage{caption}
\usepackage{listings}
\usepackage{titling}
\usepackage{chngcntr}
\counterwithin{table}{section}
\date{} 

\let\clsCenter\Center\let\clsendCenter\endCenter
\let\Center\undefined\let\endCenter\undefined
\usepackage{ragged2e}
\let\Center\clsCenter
\let\endCenter\clsendCenter

\newcommand\SentenceCase[1]{%
  \caselower{}%
  \capitalize{\thestring}%
}
\begin{singlespace}
\title{A hybrid estimation of distribution algorithm for\\
joint stratification and\\
sample allocation} 
\author
{Mervyn O'Luing,$^{1}$ Steven Prestwich,$^{1}$ S. Armagan Tarim$^{2}$
}
\end{singlespace}

\usepackage{bibentry}

\pgfplotsset{compat=1.15}
\begin{document}
\begin{singlespace}
\begin{minipage}[h]{\textwidth}
\textbf{\maketitle}
\end{minipage}
\begin{center}
    \textbf{Abstract}
\end{center}


In this study we propose a hybrid estimation of distribution algorithm (HEDA) to solve the joint stratification and sample allocation problem. This is a complex problem in which each the quality of each stratification from the set of all possible stratifications is measured its optimal sample allocation. EDAs are stochastic black-box optimization algorithms which can be used to estimate, build and sample probability models in the search for an optimal stratification. In this paper we enhance the exploitation properties of the EDA by adding a simulated annealing algorithm to make it a hybrid EDA. Results of empirical comparisons for atomic and continuous strata show that the HEDA attains the bests results found so far when compared to benchmark tests on the same data using a grouping genetic algorithm, simulated annealing algorithm or hill-climbing algorithm. However, the execution times and total execution are, in general, higher for the HEDA. 

{\let\thefootnote\relax\footnotetext{\textsuperscript{1} \textit{Insight Centre for Data Analytics, Department of Computer Science, University College Cork, Ireland.} Email: {mervyn.oluing@insight-centre.org},{ steven.prestwich@insight-centre.org }}}

{\let\thefootnote\relax\footnotetext{\textsuperscript{2} \textit{Cork University Business School, University College Cork, Ireland.} Email: {armagan.tarim@ucc.ie}}}

\end{singlespace}


\section{Introduction}

The joint determination of stratification and sample allocation designs is a complex problem in which each solution is a stratification of basic strata (either atomic or continuous). The quality of each solution is measured by the optimal sample size that can be allocated to this stratification and still meet the precision constraints set by the survey designer. This quality is evaluated by the Bethel-Chromy algorithm \citep{bethel1985optimum, chromy1987design,bethel1989sample} which is expensive (in computational terms). Previous contributions in this area includes work carried out by \citep{kozak2007modern,keskinturk2007genetic,benedetti2008tree,baillargeon2009general,baillargeon2011construction,ballin2013joint,oluing2019grouping,ballin2020optimization,o2020simulated,o2021combining}.\par
In this paper, we propose a hybrid estimation of distribution algorithm (HEDA) to solve this problem. EDAs are stochastic black-box optimization algorithms which can be used to estimate, build and sample probability models in the search for an optimal solution. We, therefore, describe the search for the lowest cost stratification as that of black-box optimization, in order to tie the problem in with existing EDA literature.
\par Rather than other methodologies where the strata have already been determined (e.g. administrative strata or the cumulative root frequency method) before evaluating the optimal sample allocation - the basic premise of this problem is that the optimal solution is unknown. 
It is a non linear problem with a rugged search space for which there are many near-optimal sample allocations (or local minima) and also perhaps more than one (i.e. attainable through multiple stratifications) optimal sample allocation (global minimum). \par  However, we cannot see the sample allocation for each stratification in advance, and thus cannot say \emph{a priori} which stratification provides the optimal allocation. We could say that we are dealing with a black box. The only sure way of determining the optimal stratification is to evaluate each solution. This is known as grid-search and is intractable for large problems. Indeed, for representative surveys, especially in official statistics, for any practical sized sampling frame the number of basic strata to be stratified (given that they may be derived from multiple auxiliary variables) can be quite large. 
\par A black-box optimization procedure explores the search space by generating solutions, evaluating them, and processing the results of this evaluation in order to generate new promising solutions \citep{gonzalez2012copulaedas}. Such procedures tend to find local minima of varying proximity to a global minimum. However they tend to be faster alternatives to grid-search. Some are deterministic, e.g. the direct search approach of \citep{hooke1961direct} or the simplex method of \citep{nelder1965simplex} and some stochastic, e.g. k-means \citep{hartwon}, grouping genetic algorithm \citep{john1975holland}, simulated annealing \citep{kirkpatrick1983optimization,vcerny1985thermodynamical} and hill-climbing \citep{lin1965computer,lin1973effective} . The latter set of procedures provide a means of attaining a solution that is "good enough" in a computing time that is "small enough" \citep{sorensen2013metaheuristics}. \par That has been our focus in earlier papers in which we presented evolutionary (a grouping genetic algorithm -\citep{oluing2019grouping}) and local search (simulated annealing algorithm - \citep{o2020simulated} algorithms and explored multi-stage combinations of clustering algorithms with a hill-climbing algorithm (\citep{o2021combining}). 
We have evaluated the performance of these algorithms by the solution quality and computation time taken to find a local minimum. In \citep{o2020simulated} and \citep{o2021combining} we also considered training times. The performance has varied according to each data stratification problem. \par To expand on the details mentioned above, we explore a new paradigm of evolutionary algorithms named estimation of distribution algorithms (EDAs) \citep{baluja1994population,muhlenbein1996recombination,larranaga2001estimation,pelikan2003hierarchical}  in the context of this problem. In particular, we consider the application of a hybridised version of an EDA and compare it against the best results achieved so far with the algorithms discussed in  \citep{oluing2019grouping}, \citep{o2020simulated} and \citep{o2021combining}. Section \ref{EDA} provides some background details on EDAs. Section \ref{HEDA2} motivates the use of a HEDA for this problem. Section \ref{joint3} summarises the objective function. \par paper. Section \ref{eval} discusses the evaluation approach. Section \ref{iris} demonstrates how the EDA component of the HEDA works. Sections \ref{emp_atomic} and \ref{emp_cont} give results of empirical comparisons of the HEDA for atomic and continuous strata. Sections \ref{conclusions2} and \ref{further2} describe our conclusions and suggestions for further work.\par

\section{Estimation of distribution algorithms (EDAs)}\label{EDA}

EDAs, which are also known as probabilistic model-building genetic algorithms (PMBGAs), belong to a family of evolutionary algorithms. They are, in effect, stochastic black-box optimization algorithms which are characterized by iteratively estimating, building and sampling probability models in the search for optimal solutions. 
\par To put it another way, EDAs work by selecting promising solutions from a population (similar to elitism in evolutionary algorithms) and building probabilistic models of those solutions. They then sample from the corresponding probability distributions to obtain new solutions \citep{lima2011model}. It is these characteristics of probabilistic model building and sampling from that model that most differentiates the EDA from other evolutionary algorithms such as the genetic algorithm.  \par 
If the model built in each generation captures the important features of selected solutions and generates new solutions with these features, then the EDA should be able to quickly converge to the optimum \citep{muhlenbein1999fda}. Indeed, assuming that the population size is large enough to ensure reliable convergence \citep{harik1999gambler,goldberg2002design}, the EDA based on the probability model provides an efficient and reliable approach to solving many optimization problems \citep{hauschild2011introduction}. \par 
Furthermore, as EDAs iteratively refine the probabilistic model using elite solutions, thereby increasing the probability of generating the global optimum, after a reasonable number of iterations, the procedure should locate the global optimum or its accurate approximation \citep{hauschild2011introduction}. Empirically, EDAs have been shown to outperform other optimisation techniques in problems such as the multi-objective knapsack problem \citep{shah2011comparative}, multi-objective monitoring network design \citep{kollat2008new} or optimising cancer chemotherapy \citep{petrovski2006optimising}. 
\par Basic EDAs use a simple probability model that has a fixed structure and learns the parameters for that model, e.g. the univariate marginal distribution algorithm (UMDA) \citep{muhlenbein1996recombination}, population-based
incremental learning (PBIL) \citep{baluja1994population} and compact genetic algorithm (cGA) \citep{harik1999compact}. Such EDAS are suitable for univariate problems. \par On the other hand, for more complex problems there are EDAs with adaptive multivariate models such as Bayesian networks (BNs) \citep{pearl1988probabilistic}, which can model complex multivariate interactions. Examples of Bayesian EDAs include the Bayesian optimization algorithm (BOA) \citep{pelikan1999boa,pelikan2003hierarchical}, the estimation of Bayesian networks algorithm (EBNA) \cite{etxeberria1999global}, and the learning factorized distribution algorithm (LFDA)\citep{muhlenbein1999fda}. \par For this problem, as we are dealing with mutually exclusive basic strata, we assume that the probability distribution of any basic stratum belonging to a particular stratum 
is independent of that for another basic stratum, and for this reason we will implement a univariate EDA based closely on the UMDA.

\section{A hybrid estimation of distribution algorithm (HEDA)}\label{HEDA2}

Even though we pointed out in section \ref{EDA} that EDAs should locate the global optimum or its near approximation in a reasonable amount of iterations, they have two main computational bottlenecks: $(1)$ fitness evaluation and $(2)$ model building. These must be addressed by efficiency enhancement techniques suitable for EDAs \citep{hauschild2011introduction}. \par The latter bottleneck we have addressed by using the simple model building approach of the UMDA - thus avoiding the computational overhead of more complex model building. The former bottleneck, clearly, is not only an issue for the EDA as it relates not only to the size of the search space but also the complexity of the problem. However, we can address this using hybridisation  \citep{hauschild2011introduction}. This is an efficiency technique the purpose of which is to speed up the performance of the EDA. 
\par In many real-world applications, evolutionary algorithms are combined with other search optimization algorithms such as local search, tabu search, simulated annealing or hill-climbing algorithms. Such metaheuristic hybrids combine the advantages of population based exploration methods (which search regions of the search space) with trajectory methods (which optimise locally and are often very successful). Simulated annealing is one example of a trajectory method which we focus on in this paper. \par Simulated annealing resembles hill-climbing in that it makes small adjustments to a candidate solution - keeping those which lead to an improvement in solution quality and gradually morphing the solution towards the closest local optimum. Hill-climbing algorithms generally attain good results especially when combined with population based search procedures \citep{lima2009loopy}.\par As an example, to solve the minimal switching graph problem  \citep{1631347} combined the exploration properties of the UMDA and the exploitation properties of the hill-climbing algorithm into a hybrid EDA to find an optimal or near-optimal solution efficiently and effectively. They compared the hybrid algorithm with the UMDA and the hill-climbing algorithm separately, and found that the performance of the hybrid EDA is significantly better than both the UMDA and the hill-climbing algorithm. \par Although, the literature on hybridisation of EDAs does not appear to have considered simulated annealing, we expect a hybrid EDA using simulated annealing to perform well. This is because hill-climbing solutions gravitate towards the nearest local minima, where they become trapped until the algorithm reaches a pre-defined end-point.  On the other hand, simulated annealing allows for a probabilistic acceptance of inferior quality solutions, enabling escape from local minima and thus attaining better quality solutions than hill-climbing. As it relates to our problem, for the reasons outlined in this section and in section \ref{EDA}, combining an EDA with simulated annealing should generate less solutions and thus translate into calling the Bethel-Chromy evaluation algorithm less often to attain optimal or near-optimal solutions.  \par  

\section{Objective function} \label{joint3}
 
A detailed consideration of the objective function is provided in \citep{oluing2019grouping}. We provide an outline below:

\begin{equation}
\begin{array}{l}
\min n = \sum_{h=1}^{H}n_{h}\\
s.t.\;\; CV\left(\hat{T}_g\right)\leq \varepsilon_g\notag \;\;\; (g=1,\ldots,G)\\
2 \leq n_{h} \leq N_{h}
\end{array}
\end{equation}

where $n$ is the sample size, $n_h$ is the sample size for stratum $h$, $N_h$ is the number of units in stratum $h$ and $H$ is the number of strata. $\hat{T}_g$ is the estimator of the population total for each one of $G$ target variable columns. The upper limit of precision $\varepsilon_g$ is expressed as a coefficient of variation $CV$ for each $\hat{T}_g$. We use the Bethel-Chromy algorithm \citep{bethel1985optimum,bethel1989sample,chromy1987design} to solve the allocation problem for a particular stratification. 

\section{Outline of the HEDA}\label{outlineHEDA}

We base our description of the hyrbid estimation of distribution aglorithm on that found in \citep{gonzalez2012copulaedas}. We start with a population of $N_{p}$ solutions either generated at random, or generated with random permutations of a starting solution (accompanying that solution). The population is evaluated and the best solutions are selected. Any selection method can be used (e.g. roulette wheel, tournament selection, Boltzmann selection, etc.) however for our algorithm we use elitism. We construct a probability model of the best solutions. We then use that model to generate new solutions to replace those not selected at the elitism stage. After a tunable number of iterations a simulated annealing algorithm is applied to a tunable number of solutions from the population (for the experiments in sections \ref{emp_atomic} and \ref{emp_cont} we apply simulated annealing to the top ranked solution). This process continues until a stopping rule has been reached. \par 

\begin{singlespace}
\begin{algorithm}[H]
\caption{Hybrid EDA}
\label{SA-EDA}
\begin{algorithmic}[1]
\State Set $i \gets 1$ and generate initial population $P_{1}$ of $N_{p}$ solutions
\State Apply simulated annealing algorithm \Comment tunable conditions 
\State Select $N_{Elite}$ promising solutions $S_{i}$ from $P_{i}$
\State Construct a probability model $M_{i}$ of the best solution from $S_{i}$.
\State Create a new population $P_{i+1}$ by sampling a set of $N_p-N_{Elite}$ new solutions according to the distribution encoded by $M_{i}$
\State set $i \gets i+1$
\State If stopping criteria (number of iterations, $I$) not met repeat steps $2$ to $6$
\end{algorithmic}
\end{algorithm}
\end{singlespace}

\section{Evaluation approach} \label{eval}
Each of the $N_p$ solutions in the population are evaluated in a two step-approach using the \emph{aggrStrata} function \citep{ballin2020optimization} and \emph{Bethel} algorithm   \citep{ballin2020optimization,de2009package}. The \emph{aggrStrata} function obtains summary statistics (including point estimates) for the basic strata. The \emph{Bethel} function uses those statistics to compute the optimal sample allocation for that stratification. 
We have coded these functions in \emph{C++} and integrated them, along with the remainder of the HEDA, into the \emph{R} software language \citep{R2021}  using the \emph{Rcpp} package \citep{Eddel} in order to speed up computation times. Accordingly, the traditional \emph{set.seed()} function in \emph{R} which was useful for reproducbility of experiments is not easily applied in \emph{Rcpp} - meaning that some further work is required before we can reproduce results exactly. 
\par Nonetheless, we found this two-step approach to be a more useful evaluation metric for the HEDA than the Bayesian Information Criterion (BIC) or total within sums of squares (TWSS) which are more suited to clustering problems and faster than the evaluation algorithm. We initially considered these as "surrogate" functions. However, solutions which attain better BIC or TWSS scores might not attain better sample allocations. This is because minimising TWSS eventually leads towards one cluster per atomic stratum whereas maximising BIC scores is useful for selecting the optimal number of clusters, but not necessarily the optimal arrangement of items within clusters. \par

\section{Example of EDA component of the HEDA on the iris data set}\label{iris}

To demonstrate how an estimation of distribution algorithm works we use the iris data set \citep{anderson,fisher,R2021} and atomic strata. We select Petal Length and Petal Width as the target variables. We use Sepal Length and Species as auxiliary variables. We convert Sepal Length to a categorical variable with 3 bins using the k-means algorithm \citep{hartwon} and a seed of $1234$. We use an upper coefficient of variation level of $0.05$ for the target variables.  The cross product of the categorical version of Sepal Length with Species results in 8 atomic strata. We group these 8 atomic strata into $H$ strata, in this example $H=3$ or $H=4$, i.e. some of the solutions have three strata and some have four.

\begin{table}[H]
\centering
\tiny
\caption{Summary statistics for the $L$ atomic strata}
\label{iris_sum}
\begin{tabular}{|l|l|l|l|l|l|}
\hline
\textbf{Atomic Stratum, $l$} & \textbf{N} & \textbf{M1} & \textbf{M2} & \textbf{S1} & \textbf{S2} \\ \hline
\textbf{a}              & 40         & 1.46        & 0.24        & 0.17        & 0.10        \\ \hline
\textbf{b}              & 5          & 3.40        & 1.10        & 0.30        & 0.15        \\ \hline
\textbf{c}              & 1          & 4.50        & 1.70        & 0.00        & 0.00        \\ \hline
\textbf{d}              & 10         & 1.48        & 0.29        & 0.17        & 0.09        \\ \hline
\textbf{e}              & 31         & 4.23        & 1.31        & 0.36        & 0.19        \\ \hline
\textbf{f}              & 12         & 5.07        & 1.88        & 0.22        & 0.27        \\ \hline
\textbf{g}              & 14         & 4.64        & 1.45        & 0.21        & 0.11        \\ \hline
\textbf{h}              & 37         & 5.74        & 2.08        & 0.50        & 0.25        \\ \hline
\end{tabular}
\end{table}

We initialise a population size of $N_{p}$ solutions (in this case $N_{p}=5$) each of size $L$, i.e. the number of atomic strata, where $l = 1, 2, \dots , L$.  As can be seen in table \ref{iris_sum} we initially use letters of the alphabet to differentiate between atomic strata, however in subsequent tables an integer denotes to which one of the $H$ strata each atomic stratum belongs. 
Each solution is evaluated and ranked.

\begin{table}[H]
\centering
\tiny
\caption{Population of size $N_p$ with solution quality and rank}
\begin{tabular}{|l|l|l|l|l|l|l|l|l|l|}
\hline
\multicolumn{8}{|c|}{\textbf{Initial Population}} & \textbf{Quality} & \textbf{Rank} \\ \hline
3    & 2    & 2    & 3    & 2   & 1   & 2   & 2   & 19.19            & 3             \\ \hline
3    & 2    & 4    & 3    & 2   & 1   & 2   & 3   & 46.77            & 5             \\ \hline
3    & 2    & 4    & 1    & 2   & 1   & 2   & 4   & 14.48            & 2             \\ \hline
3    & 2    & 4    & 3    & 2   & 1   & 1   & 4   & 10.65            & 1             \\ \hline
3    & 3    & 2    & 3    & 2   & 1   & 2   & 2   & 30.30            & 4             \\ \hline
\end{tabular}
\end{table}

We create a new selected population of the best solutions (in this example we select 2 solutions) in a process that is analgous with elitism in evolutionary algorithms. 

\begin{table}[H]
\centering
\tiny
\caption{Selected population of best elite solutions}
\begin{tabular}{|l|l|l|l|l|l|l|l|}
\hline
\multicolumn{8}{|c|}{\textbf{Selected Population}} \\ \hline
3    & 2    & 4    & 3    & 2    & 1   & 1   & 4   \\ \hline
3    & 2    & 4    & 1    & 2    & 1   & 2   & 4   \\ \hline
\end{tabular}
\end{table}

We then construct a probabilistic model of the solutions in the selected population with the aim of estimating the probability distribution for each of the $H$ strata for each atomic stratum. The probability of a candidate solution (or vector, $V$) where ($V = V_{1}, V_{2}, ..., V_{L}$) is the product of probabilities of individual strata for each atomic stratum:
\begin{equation}
  p(V_{1}, V_{2}, ..., V_{L}) =p(V_{1})\times p(V_{2}), ..., \times p(V_{L})  
\end{equation}

where $p(V_{l})$ is the probability of variable $(V_{l}= v_{l})$, $v_l$ is an integer between $1$ and $H$ and $p(V_{1}, V_{2}, ..., V_{L})$ is the probability of the candidate solution $(V_{1}, V_{2}, ..., V_{L})$ and $H$ represents the number of strata which are labelled individually by an integer between $1$ and $H$. The univariate model for the  $L$ variables consists of $L$ vectors containing probabilities of an atomic stratum belonging to the different strata. The probabilities different strata sum to 1. 

The joint probability distribution is factorized as a product of independent univariate marginal distributions, which are estimated from
marginal frequencies \citep{paul2002linear}:
\begin{equation}
    p(V_l) = \frac{\sum_{j=1}^{N_{Elite}} \delta_{j} (V_{l}=v_{l} | N_{Elite}^{iter-1})}{N_{Elite}^{iter-1}}
\end{equation}

where $N_{Elite}^{iter-1}$is the number of elite solutions from the previous iteration, $\delta_{j} (V_{l}=v_{l} | N_{Elite}^{iter-1})=1$ if $V_{l}=v_{l}$ in the $j^{th}$case of $N_{Elite}^{iter-1}$, and $0$ otherwise.

\begin{table}[H]
\tiny
\centering
\caption{Model estimating the probability of each atomic stratum $l$ belonging to each stratum $h$}
\begin{tabular}{|l|l|l|l|l|l|l|l|l|}
\hline
\textbf{Stratum, $h$} & \multicolumn{8}{c|}{\textbf{Probability Model }}                                                       \\ \hline
1                & 0          & 0          & 0          & 0.5        & 0          & 1          & 0.5        & 0          \\ \hline
2                & 0          & 1          & 0          & 0          & 1          & 0          & 0.5        & 0          \\ \hline
3                & 1          & 0          & 0          & 0.5        & 0          & 0          & 0          & 0          \\ \hline
4                & 0          & 0          & 1          & 0          & 0          & 0          & 0          & 1          \\ \hline
\textbf{Total}   & \textbf{1} & \textbf{1} & \textbf{1} & \textbf{1} & \textbf{1} & \textbf{1} & \textbf{1} & \textbf{1} \\ \hline
\end{tabular}
\end{table}

We generate new solutions from this model to replace the non-elite solutions by sampling from that distribution. We evaluate the new solutions and rank all solutions by their quality. This offspring population returns a solution quality of $9.34$ which is the global minimum (to find the global minimum we evaluated each of the 4,140 possible partitions of the 8 atomic strata).

\begin{table}[H]
\centering
\tiny
\caption{Offspring population with solution quality and rank}
\begin{tabular}{|l|l|l|l|l|l|l|l|l|l|}
\hline
\multicolumn{8}{|c|}{\textbf{Offspring Population}} & \textbf{Quality} & \textbf{Rank} \\ \hline
3    & 2    & 4    & 3    & 2    & 1    & 1   & 4   & 10.65            & 2             \\ \hline
3    & 2    & 4    & 1    & 2    & 1    & 2   & 4   & 14.48            & 3             \\ \hline
3    & 2    & 4    & 3    & 2    & 1    & 1   & 4   & 10.65            & 2             \\ \hline
3    & 2    & 4    & 3    & 2    & 1    & 2   & 4   & 9.34             & 1             \\ \hline
3    & 2    & 4    & 1    & 2    & 1    & 2   & 4   & 14.48            & 3             \\ \hline
\end{tabular}
\end{table}

Normally the algorithm repeats the above process until a stopping criterion has been met. \par The extreme probabilities — $0$ and $1$ — mean that unless some mitigating measure is introduced the atomic stratum $l$ would remain fixed forever to stratum $h$ at a probability of either $0$ or $1$, obstructing some regions of the search space \citep{dang2019level}. However, at advanced stages of the search, e.g. when using a starting solution, the stratifications are of good quality and atomic stratum may already be in the correct stratum (i.e. the stratum it would be allocated to in the optimal solution).  \par In the above experiment it is clear that for all solutions in the population there are a number of cases where there is only one stratum choice available for an atomic stratum (i.e. it is assigned to that stratum with a probability of $1$). Here we are very close to a global minimum and we get there on the second step  - without adjusting the probability of strata assignments. On a more general level however, the hybrid EDA uses mutation and add strata components (see \citep{oluing2019grouping}), as well as  simulated annealing (see \citep{o2020simulated} which enable escape from this occurrence. The UMDA also has alternative approaches of avoiding $1,0$ probabilities (such as capping the probability interval to between $\frac{1}{L}$ and $(1-\frac{1}{L})$) e.g. \citep{doerr2020univariate}, however, for our purposes they are not required. 

\section{Empirical comparisons for atomic strata}\label{emp_atomic} 

\subsection{Background details}\label{backgrounddetails}

We compare the performance of the HEDA with the best solution qualities from the experiments we carried out on the same data sets, with the grouping genetic algorithm, simulated annealing algorithm and combinations of clustering algorithms with hill-climbing. We use the results from these experiments as a benchmark for comparing the HEDA. More details on these experiments are provided in \cite{oluing2019grouping,o2020simulated,o2021combining}.

The data sets including target and auxiliary variables are summarised in table \ref{variables3}. We explain the meaning of each variable in table \ref{breakdown} in section \ref{appendix}.

\begin{table}[H]
\caption{Summary by data set of the variables, number of records and atomic strata}
\label{variables3}
\tiny
\centering
\begin{tabular}{|l|l|l|l|l|}
\hline
\textbf{Data set} & \textbf{\begin{tabular}[c]{@{}l@{}}Target \\Variables\end{tabular}} & \textbf{\begin{tabular}[c]{@{}l@{}}Auxiliary Variables\end{tabular}} & \textbf{\begin{tabular}[c]{@{}l@{}}Number of  \\Records\end{tabular}} & \textbf{\begin{tabular}[c]{@{}l@{}}Number of \\Atomic \\Strata\end{tabular}} \\ \hline
\textbf{Swiss Municipalities} & Surfacesbois, Airbat & POPTOT , HApoly & 2,896 & 579 \\ \hline
\textbf{\begin{tabular}[c]{@{}l@{}}American Community  \\ Survey, 2015\end{tabular}} & \begin{tabular}[c]{@{}l@{}}HINCP, VALP,   \\ SMOCP, INSP\end{tabular} & \begin{tabular}[c]{@{}l@{}}BLD, TEN, WKEXREL,   \\ WORKSTAT, HFL, YBL\end{tabular} & 619,747 & 123,007 \\ \hline
\textbf{Kiva Loans} & \begin{tabular}[c]{@{}l@{}}terminmonths,  \\ lendercount,   \\ loanamount\end{tabular} & \begin{tabular}[c]{@{}l@{}}sector, currency, activity,   \\ region, partnerid\end{tabular} & 614,361 & 84,897 \\ \hline
\end{tabular}
\end{table}

Table \ref{prev_res} provides details of the best sample sizes and the corresponding algorithms from the aforementioned which they were attained with.  

\begin{table}[H]
\caption{Summary by data set of the sample size and evaluation time for the final stage in multi stage combinations of clustering algorithms with the simulated annealing algorithm and hill-climbing algorithm}
\label{prev_res}
\centering
\tiny
\begin{tabular}{|l|l|l|l|l|l|}
\hline
\textbf{Dataset} & \textbf{\begin{tabular}[c]{@{}l@{}}Initial \\Stage(s)\end{tabular}} & \textbf{\begin{tabular}[c]{@{}l@{}}Final \\Stage\end{tabular}} & \textbf{\begin{tabular}[c]{@{}l@{}}Sample \\Size\end{tabular}} & \textbf{\begin{tabular}[c]{@{}l@{}}Execution \\Time\end{tabular}} & \textbf{\begin{tabular}[c]{@{}l@{}}Total \\Execution \\Time\end{tabular}} \\ \hline
\textbf{Swiss Municipalities} & Kmeans & SAA & 125.17 & 248.91 & 8,808.63 \\ \hline
\textbf{\begin{tabular}[c]{@{}l@{}}American Community \\ Survey, 2015\end{tabular}} & SOM, FC & HC & 9,065.45 & 6,298.81 & 6,298.81 \\ \hline
\textbf{Kiva Loans} & FC & HC & 6,326.35 & 6,018.61 & 6,018.61 \\ \hline
\end{tabular}
\end{table}

However, as we have sought to develop algorithms that provide solutions that are good enough in a computing time that is small enough we also report evaluation (execution) and training (total execution) times to provide context to the results. In table \ref{prev_res}, \emph{SAA} refers to simulated annealing algorithm, \emph{SOM} applies to self-organising map, \emph{FC} relates to Fuzzy Clustering and \emph{HC} indicates hill-climbing. \par For comparison purposes execution time and total execution time relates to the final stage algorithms only. Note: there was no fine-tuning required for the hill-climbing algorithm. Details of the fine-tuning time for the initial stage solutions are available in \citep{o2020simulated} and \citep{o2021combining}. 

\subsection{Hyperparameters}

The hyperparameters used in the experiments for the HEDA are provided in Table \ref{atomic_hyper}.

\begin{table}[H]
\tiny
\centering
\caption{Hyperparameters for the hybrid estimation of distribution algorithm (HEDA)}
\label{atomic_hyper}
\begin{tabular}{|l|l|l|l|l|l|l|}
\hline
\textbf{Dataset} &
  \textbf{Iterations, I} &
  \textbf{\begin{tabular}[c]{@{}l@{}}SAA \\ Frequency\end{tabular}} &
  \textbf{\begin{tabular}[c]{@{}l@{}}Mutation \\ Chance\end{tabular}} &
  \textbf{\begin{tabular}[c]{@{}l@{}}Elitism \\ Rate\end{tabular}} &
  \textbf{\begin{tabular}[c]{@{}l@{}}Add \\ Strata \\ Factor\end{tabular}} &
  \textbf{\begin{tabular}[c]{@{}l@{}}Temperature, \\ T\end{tabular}} \\ \hline
Swiss Municipalities            & 3,000       & 100 & 0.000185221 & 0.20 & 0.000000146 & 0.000000027 \\ \hline
American Community Survey, 2015 & 200         & 10  & 0.000000016 & 0.30 & 0.000000054 & 0.000655857 \\ \hline
Kiva Loans                      & 600         & 100 & 0.000038838 & 0.30 & 0.000000739 & 0.001400962 \\ \hline
\textbf{Dataset} &
  \textbf{\begin{tabular}[c]{@{}l@{}}Decrement \\ Constant,DC\end{tabular}} &
  \textbf{\begin{tabular}[c]{@{}l@{}}Max \\ number of \\ sequences,\\ maxit\end{tabular}} &
  \textbf{\begin{tabular}[c]{@{}l@{}}Length \\ of \\ sequence, \\ J\end{tabular}} &
  \textbf{\begin{tabular}[c]{@{}l@{}}\% of L \\ for maximum \\ q value, \\ Lmax\%\end{tabular}} &
  \textbf{\begin{tabular}[c]{@{}l@{}}Probability \\ of New \\ Stratum, \\ P(H + 1)\end{tabular}} &
  \textbf{\begin{tabular}[c]{@{}l@{}}Population Size,\\ $N_p$\end{tabular}} \\ \hline
Swiss Municipalities            & 0.941411680 & 2   & 20,000      & 0.10 & 0.009183601 & 20          \\ \hline
American Community Survey, 2015 & 0.815042754 & 20  & 5,000       & 0.01 & 0.000000609 & 20          \\ \hline
Kiva Loans                      & 0.954218273 & 2   & 10,000      & 0.02 & 0.000052373 & 20          \\ \hline
\end{tabular}
\end{table}

The hyperparameters operate in this way, \emph{Iterations, $I$} indicates the number of iterations the algorithm runs for. The algorithm stops at the last iteration. The HEDA builds and samples from a \emph{probability model, $M_{i}$} once every iteration. The simulated annealing algorithm also runs after a tunable number of iterations, i.e. for the Swiss Municipalities and Kiva Loans experiments it runs once every $100$ iterations whereas for the American Community Survey experiment it runs once every $10$ iterations. The simulated annealing algorithm runs each time for a maximum number of sequences with the length of each sequence varying for each experiment. See \citep{o2020simulated} for more details on the hyperparameters for the simulated annealing algorithm. 
\par The following hyperparameters are described in more detail in \citep{oluing2019grouping}. However in brief, \emph{Elitism rate} indicates the proportion of the ranked population that is carried over to the next iteration, \emph{Mutation chance} indicates the probability of each atomic stratum being moved to another stratum every iteration. Similarly, \emph{Add Strata Factor} is the probability of an atomic stratum moving to a new stratum. \emph{Population size, $N_p$} is the number of solutions used for exploration and evolution each iteration. \par The hyperparameters often need to be fine-tuned to suit the characteristics of a particular data set. We fine-tuned the hyperparameters for these experiments using the methodology on as described by \citep{bischla2017mlrmbo} in the R software language \citep{R2021}. Finally, although we have stated the hyperparameters, the solution quality attained with them in the HEDA may not be reproduced exactly on their first application. See table \ref{HEDAhyper_atomic} for details of the hyperaparameters, execution and total execution times in the training process.  \par 

\subsection{Results}

The results of these experiments are provided in Table \ref{atomic_results}.

\begin{table}[H]
\tiny
\centering
\caption{Summary by data set of the benchmark sample size and evaluation time for the multi stage combinations of clustering algorithms with the HEDA}
\label{atomic_results}
\begin{tabular}{|l|l|l|l|l|l|}
\hline
\textbf{Data set} & \textbf{Initial Stage(s)} & \textbf{Final Stage} & \textbf{Sample Size} & \textbf{Execution Time} & \textbf{Total Execution Time} \\ \hline
\textbf{Swiss Municipalities}            & Kmeans     & HEDA & 122.39   & 1,316.11  & 47,045.87  \\ \hline
\textbf{American Community Survey, 2015} & SOM and FC & HEDA & 8,800.53 & 28,857.52 & 550,676.58 \\ \hline
\textbf{Kiva Loans}                      & Kmeans     & HEDA & 6,246.74 & 3,292.97  & 53,006.37  \\ \hline
\end{tabular}
\end{table}

We compare the sample sizes attained in table \ref{prev_res} and table \ref{atomic_results} and present the results below in table \ref{res_comp}. 

\begin{table}[H]
\tiny
\centering
\caption{Ratio comparison of the sample sizes for the benchmark results and the HEDA results}
\label{res_comp}
\begin{tabular}{|l|l|l|l|}
\hline
\textbf{Data set} & \textbf{Benchmark results} & \textbf{HEDA Results} & \textbf{Ratio comparisons} \\ \hline
\textbf{Swiss Municipalities}            & 125.17   & 122.39   & 0.98 \\ \hline
\textbf{American Community Survey, 2015} & 9,065.45 & 8,800.53 & 0.97 \\ \hline
\textbf{Kiva Loans}                      & 6,326.35 & 6,246.74 & 0.99 \\ \hline
\end{tabular}
\end{table}
The ratio of the HEDA results to the benchmark results indicate that, although they are similar, the HEDA results are of better quality in all cases. However, as our focus has been on obtaining results that are good enough in a time that is small enough these results should be considered in the context of execution and total execution times. \par
Table \ref{times1} compares the execution times for the HEDA and those for the final stage algorithm in the benchmark results.

\begin{table}[H]
\tiny\centering
\caption{Ratio comparison of the execution times and total
execution times for the benchmark results and the HEDA results}
\label{times1}
\begin{tabular}{|l|l|l|l|l|l|l|}
\hline
 &
  \textbf{\begin{tabular}[c]{@{}l@{}}Benchmark \\ Results\end{tabular}} &
  \textbf{\begin{tabular}[c]{@{}l@{}}HEDA \\ Results\end{tabular}} &
   &
  \textbf{\begin{tabular}[c]{@{}l@{}}Benchmark \\ Results\end{tabular}} &
  \textbf{\begin{tabular}[c]{@{}l@{}}HEDA \\ Results\end{tabular}} &
   \\ \hline
\textbf{Data set} &
  \textbf{\begin{tabular}[c]{@{}l@{}}Execution \\ Time\end{tabular}} &
  \textbf{\begin{tabular}[c]{@{}l@{}}Execution \\ Time\end{tabular}} &
  \textbf{\begin{tabular}[c]{@{}l@{}}Ratio \\ Comparisons\end{tabular}} &
  \textbf{\begin{tabular}[c]{@{}l@{}}Total \\ Execution \\ Time\end{tabular}} &
  \textbf{\begin{tabular}[c]{@{}l@{}}Total \\ Execution \\ Time\end{tabular}} &
  \textbf{\begin{tabular}[c]{@{}l@{}}Ratio \\ Comparisons\end{tabular}} \\ \hline
\textbf{Swiss Municipalities}            & 248.91   & 1,316.11  & 5.29 & 8,808.63 & 47,045.87  & 5.34  \\ \hline
\textbf{American Community Survey, 2015} & 6,298.81 & 28,857.52 & 4.58 & 6,298.81 & 550,676.58 & 87.43 \\ \hline
\textbf{Kiva Loans}                      & 6,018.61 & 3,292.97  & 0.55 & 6,018.61 & 53,006.37  & 8.81  \\ \hline
\end{tabular}
\end{table}

In brief the results indicate that overall, for these experiments, the HEDA takes longer to execute and fine-tune than the SAA for the Swiss Municipalities,or the combination of clustering algorithms with hill-climbing for the American Community Survey, 2015 and Kiva Loans experiments. 




\section{Empirical comparisons for continuous strata}\label{emp_cont} 

Using the same methodology applied in section \ref{emp_atomic} we ran experiments using the same data sets comparing the performance of the HEDA on the stratification of continuous strata with the benchmark results from experiments we described in \citep{oluing2019grouping}, \citep{o2020simulated} and \citep{o2021combining}. As before we first provide a summary of the target and auxiliary variables in table \ref{xandy}. 

\begin{table}[H]
\tiny
\centering
\caption{Target and auxiliary variables for the Continuous method}
\label{xandy}
\begin{tabular}{|l|l|l|l|l|}
\hline
\textbf{Data set} &
  \multicolumn{4}{l|}{\textbf{\begin{tabular}[c]{@{}l@{}}Target variables \\ Auxiliary variables\end{tabular}}} \\ \hline
\textbf{Swiss Municipalities} & Airbat           & Surfacesbois  &              &  \\ \hline
\textbf{\begin{tabular}[c]{@{}l@{}}American Community \\ Survey, 2015\end{tabular}} &
  HINCP &
  VALP &
  SMOCP &
  INSP \\ \hline
  \textbf{US Census, 2000}           & HHINCOME         &               &              &  \\ \hline
\textbf{Kiva Loans}          & term\_in\_months & lender\_count & loan\_amount &  \\ \hline
\textbf{\begin{tabular}[c]{@{}l@{}}UN Commodity Trade \\ Statistics data\end{tabular}} &
  trade\_usd &
   &
   &
   \\ \hline
\end{tabular}
\end{table}

Likewise table \ref{prev_res_cont} provides details of the best sample sizes and the corresponding algorithms from the aforementioned which they were attained with.  

\begin{table}[H]
\tiny
\centering
\caption{Summary by data set of the sample size and evaluation time for the final stage of multi stage combinations of clustering algorithms with the hill-climbing algorithm}
\label{prev_res_cont}
\begin{tabular}{|l|l|l|l|l|l|}
\hline
\textbf{Dataset} & \textbf{Initial Stage(s)} & \textbf{Final Stage} & \textbf{Sample Size} & \textbf{Execution Time} & \textbf{Total Execution Time} \\ \hline
\textbf{Swiss Municipalities} & NG and EM & HC & 110 & 38.61 & 38.61 \\ \hline
\textbf{\begin{tabular}[c]{@{}l@{}}American Community \\ Survey, 2015\end{tabular}} & NG and FC & HC & 2,753.55 & 28,632.09 & 28,632.09 \\ \hline
\textbf{Kiva Loans} & EM & HC & 1,752.82 & 7,370.53 & 7,370.53 \\ \hline
\end{tabular}
\end{table}

\subsection{Hyperparameters}

The hyperparameters used in the HEDA are provided below in Table \ref{cont_hyper}.

\begin{table}[H]
\tiny
\centering
\caption{Hyperparameters for the hybrid estimation of distribution algorithm (HEDA)}
\label{cont_hyper}
\begin{tabular}{|l|l|l|l|l|l|l|}
\hline
\textbf{Dataset} &
  \textbf{\begin{tabular}[c]{@{}l@{}}Iterations, \\ I\end{tabular}} &
  \textbf{\begin{tabular}[c]{@{}l@{}}SAA \\ Frequency\end{tabular}} &
  \textbf{\begin{tabular}[c]{@{}l@{}}Mutation \\ Chance\end{tabular}} &
  \textbf{\begin{tabular}[c]{@{}l@{}}Elitism \\ Rate\end{tabular}} &
  \textbf{\begin{tabular}[c]{@{}l@{}}Add \\ Strata \\ Factor\end{tabular}} &
  \textbf{\begin{tabular}[c]{@{}l@{}}Temperature, \\ T\end{tabular}} \\ \hline
Swiss Municipalities            & 1,000     & 100 & 0.0000008 & 0.2       & 0.0000000 & 0.0000058 \\ \hline
American Community Survey, 2015 & 30        & 10  & 0.0000003 & 0.3       & 0.0000010 & 0.0000010 \\ \hline
Kiva Loans                      & 90        & 10  & 0.0000004 & 0.3       & 0.0000001 & 0.0000001 \\ \hline
\textbf{Dataset} &
  \textbf{\begin{tabular}[c]{@{}l@{}}Decrement \\ Constant,\\ DC\end{tabular}} &
  \textbf{\begin{tabular}[c]{@{}l@{}}Max \\ number \\ of \\ sequences,\\ maxit\end{tabular}} &
  \textbf{\begin{tabular}[c]{@{}l@{}}Length \\ of \\ sequence, \\ J\end{tabular}} &
  \textbf{\begin{tabular}[c]{@{}l@{}}\% of L \\ for \\ maximum \\ q value, \\ Lmax\%\end{tabular}} &
  \textbf{\begin{tabular}[c]{@{}l@{}}Probability \\ of \\ New \\ Stratum, \\ P(H + 1)\end{tabular}} &
  \textbf{\begin{tabular}[c]{@{}l@{}}Population \\ Size,\\ N\_p\end{tabular}} \\ \hline
Swiss Municipalities            & 0.6266293 & 10  & 3,000     & 0.0195154 & 0.0000377 & 20        \\ \hline
American Community Survey, 2015 & 0.9849021 & 2   & 50,000    & 0.0113557 & 0.0000001 & 20        \\ \hline
Kiva Loans                      & 0.9844605 & 2   & 20,000    & 0.0156086 & 0.0000008 & 20        \\ \hline
\end{tabular}
\end{table}

Table \ref{HEDAhyper_cont} outlines the hyperaparameters used, along with the execution and total execution times resulting from the training process.  

\subsection{Results}

The results of the HEDA experiments on continuous strata are provided in Table \ref{cont_results}.

\begin{table}[H]
\centering
\tiny
\caption{Summary by data set of the sample size and evaluation time for the final stage in multi stage combinations of clustering algorithms with the HEDA}
\label{cont_results}
\begin{tabular}{|l|l|l|l|l|l|}
\hline
\textbf{Data set} & \textbf{Initial Stage(s)} & \textbf{Final Stage} & \textbf{Sample Size} & \textbf{Execution Time} & \textbf{Total Execution Time} \\ \hline
\textbf{Swiss Municipalities}            & Kmeans        & HEDA & 104.93   & 207.11    & 6,792.40   \\ \hline
\textbf{American Community Survey, 2015} & NG and FC     & HEDA & 2,619.06 & 22,663.52 & 247,314.63 \\ \hline
\textbf{Kiva Loans}                      & NG and Kmeans & HEDA & 1,576.80 & 8,791.12  & 88,951.97  \\ \hline
\end{tabular}
\end{table}

As before, we compare the sample sizes attained in table \ref{prev_res_cont} and table \ref{cont_results} and present the results below in table \ref{res_comp_cont}. 

\begin{table}[H]
\tiny
\centering
\caption{Ratio comparison of the sample sizes for the benchmark results and the HEDA results}
\label{res_comp_cont}
\begin{tabular}{|l|l|l|l|}
\hline
\textbf{Data set}                        & \textbf{Benchmark Results} & \textbf{HEDA Results} & \textbf{Ratio Comparisons} \\ \hline
\textbf{Swiss Municipalities} & 110      & 104.930419822273 & 0.95 \\ \hline
\textbf{American Community Survey, 2015} & 2,753.55                   & 2619.06326917666      & 0.95                       \\ \hline
\textbf{Kiva Loans}           & 1,752.82 & 1,576.80         & 0.90 \\ \hline
\end{tabular}
\end{table}

The ratio of the HEDA results to the benchmark results indicate that, the HEDA results are of better quality in all three experiments.  \par 
Table \ref{times_cont} indicates that the execution times and total execution times are, overall, significantly greater for the HEDA.  

\begin{table}[H]
\tiny
\centering
\caption{Ratio comparison of the execution times and total execution times for the benchmark results and the HEDA results }
\label{times_cont}
\begin{tabular}{|l|l|l|l|l|l|l|}
\hline
 &
  \textbf{\begin{tabular}[c]{@{}l@{}}Benchmark \\ Results\end{tabular}} &
  \textbf{\begin{tabular}[c]{@{}l@{}}HEDA \\ Results\end{tabular}} &
   &
  \textbf{\begin{tabular}[c]{@{}l@{}}Benchmark \\ Results\end{tabular}} &
  \textbf{\begin{tabular}[c]{@{}l@{}}HEDA \\ Results\end{tabular}} &
   \\ \hline
\textbf{Data set} &
  \textbf{\begin{tabular}[c]{@{}l@{}}Execution \\ Time\end{tabular}} &
  \textbf{\begin{tabular}[c]{@{}l@{}}Execution \\ Time\end{tabular}} &
  \textbf{\begin{tabular}[c]{@{}l@{}}Ratio \\ Comparisons\end{tabular}} &
  \textbf{\begin{tabular}[c]{@{}l@{}}Total \\ Execution \\ Time\end{tabular}} &
  \textbf{\begin{tabular}[c]{@{}l@{}}Total \\ Execution \\ Time\end{tabular}} &
  \textbf{\begin{tabular}[c]{@{}l@{}}Ratio \\ Comparisons\end{tabular}} \\ \hline
\textbf{Swiss Municipalities}            & 38.61     & 207.11    & 5.36 & 95.33     & 6,792.40   & 71.25 \\ \hline
\textbf{American Community Survey, 2015} & 28,632.09 & 22,663.52 & 0.79 & 41,740.96 & 247,314.63 & 5.92  \\ \hline
\textbf{Kiva Loans}                      & 7,370.53  & 8,791.12  & 1.19 & 7,428.64  & 88,951.97  & 11.97 \\ \hline
\end{tabular}
\end{table}

\section{Conclusions} \label{conclusions2}

Although these experiments are intended to be demonstrative rather than exhaustive, the multi-stage combination of an initial solution with the HEDA generally attains better sample sizes than the combinations of initial solutions with either the SAA or HC algorithm which had attained the best results in earlier experiments (that also included the GGA). 
\par The execution times and total execution are, generally, significantly higher for the HEDA than some of the best alternative performing algorithms (especially the clustering algorithms considered in \citep{o2021combining}). However, it is worth recalling that the HC algorithm in the above experiments had converged on a solution and would not have attained better results without adjusting the stopping conditions for algorithm or adjusting the solution. We should also bear in mind that the HEDA has found the best results so far for these experiments, and the execution and total times for the benchmark algorithms to find solutions of that quality (or better), using either the same or alternative starting solution, have not been evaluated. \par That said, the HEDA is useful for a survey designer with expert knowledge of the sampling frame and the stratification of basic strata (atomic and continuous)- for whom the requirement to fine-tune hyperparameters is mitigated by prior knowledge. It is also useful for survey designers for whom the fine-tuning times can be included in their computation budget. \par
Nonetheless, because of the combination of probability models, explorative and exploitative search we feel the HEDA is well placed to generally outperform when it comes to attaining results which are good enough in a time that is small enough.

\section{Further Work} \label{further2}

in the case of the HEDA adjustments can be made in order to use a seed function for reproducibility of results. Optimised versions of the GGA, SAA and basic EDA can also be implemented in \emph{Rcpp}. It might be useful to extend the functionality of the HEDA to the stratification approach used for spatial sampling by \cite{ballin2020optimization}. Furthermore, as we kept the population size, and the number of solutions to which we applied simulated annealing to every iteration, the same for all experiments, there is scope to fine those hyperparameters. Lastly, we could investigate more complex techniques such as the Bayesian optimisation algorithm, the estimation of Bayesian networks algorithm, the learning factorized distribution algorithm or mutual information maximising input clustering (MIMIC) \citep{de1997mimic} on this problem.

\section{Acknowledgements}
We acknowledge the useful discussions with Dr. Thierry Vallée of the Central Statistics Office in general with respect to the progress of the PhD, but also with respect to tweaking the \emph{Rcpp} code outlined above. \par This material is based upon work supported by the Insight Centre for Data Analytics and Science Foundation Ireland under Grant No. 12/RC/2289-P2 which is co-funded under the European Regional Development Fund. Also, this publication has emanated from research supported in part by a grant from Science Foundation Ireland under Grant number 16/RC/3918 which is co-funded under the European Regional Development Fund.

\bibliographystyle{chicago}

\Urlmuskip=0mu plus 1mu\relax
\def\UrlBreaks{\do\/\do-}

\newpage
 \chapter{Appendix} 

\section{Descriptions of target and auxiliary variables for the HEDA experiment data sets}\label{appendix}

\begin{table}[H]
\centering
\tiny
\caption{Breakdown by data set of descriptions for target and auxiliary variables}\label{breakdown}
\begin{tabular}{|l|l|l|}
\hline
\textbf{Data set}                         & \textbf{Variable} & \textbf{Description}                                          \\ \hline
\textbf{SwissMunicipalities}             & Surfacebois       & {\color[HTML]{374151} wood area.}                             \\ \hline
\textbf{SwissMunicipalities}             & Airbat            & {\color[HTML]{374151} area with buildings.}                   \\ \hline
\textbf{SwissMunicipalities}             & POPTOT            & {\color[HTML]{374151} total population}                       \\ \hline
\textbf{SwissMunicipalities}             & Hapoly            & {\color[HTML]{374151} municipality area.}                     \\ \hline
\textbf{American Community Survey, 2015} & HINCP             & Household income (past 12 months)                             \\ \hline
\textbf{American Community Survey, 2015} & VALP              & Property value                                                \\ \hline
\textbf{American Community Survey, 2015} & SMOCP             & Selected monthly owner costs                                  \\
\hline
\end{tabular}
\end{table}

\section{Fine-tuning the hyperparameters for the hybrid Estimation of Distribution Algorithm - Atomic Strata}\label{appendixatom}
{
\tiny
\centering
\begin{longtable}{|llllllll|}
\caption{Hyperparameters for the HEDA in the case of Atomic Strata}
\label{HEDAhyper_atomic}\\
\hline
\multicolumn{8}{|c|}{\textbf{Swiss Municipalities}} \\ \hline
\endfirsthead
\endhead
\multicolumn{1}{|l|}{\textbf{\begin{tabular}[c]{@{}l@{}}Fine-tuning \\ iteration \\ No.\end{tabular}}} &
  \multicolumn{1}{l|}{\textbf{Iterations, I}} &
  \multicolumn{1}{l|}{\textbf{\begin{tabular}[c]{@{}l@{}}SAA \\ Frequency\end{tabular}}} &
  \multicolumn{1}{l|}{\textbf{\begin{tabular}[c]{@{}l@{}}Mutation \\ Chance\end{tabular}}} &
  \multicolumn{1}{l|}{\textbf{\begin{tabular}[c]{@{}l@{}}Elitism \\ Rate\end{tabular}}} &
  \multicolumn{1}{l|}{\textbf{\begin{tabular}[c]{@{}l@{}}Add \\ Strata \\ Factor\end{tabular}}} &
  \multicolumn{1}{l|}{\textbf{\begin{tabular}[c]{@{}l@{}}Temperature, \\ T\end{tabular}}} &
  \textbf{\begin{tabular}[c]{@{}l@{}}Decrement Constant,\\ DC\end{tabular}} \\ \hline
\multicolumn{1}{|l|}{1} &
  \multicolumn{1}{l|}{5,000} &
  \multicolumn{1}{l|}{100} &
  \multicolumn{1}{l|}{0.0004011} &
  \multicolumn{1}{l|}{0.15} &
  \multicolumn{1}{l|}{0.0000008} &
  \multicolumn{1}{l|}{0.00000002} &
  0.6933468 \\ \hline
\multicolumn{1}{|l|}{2} &
  \multicolumn{1}{l|}{3,000} &
  \multicolumn{1}{l|}{100} &
  \multicolumn{1}{l|}{0.0009943} &
  \multicolumn{1}{l|}{0.15} &
  \multicolumn{1}{l|}{0.0000002} &
  \multicolumn{1}{l|}{0.00000005} &
  0.5736595 \\ \hline
\multicolumn{1}{|l|}{3} &
  \multicolumn{1}{l|}{5,000} &
  \multicolumn{1}{l|}{100} &
  \multicolumn{1}{l|}{0.0002985} &
  \multicolumn{1}{l|}{0.2} &
  \multicolumn{1}{l|}{0.0000005} &
  \multicolumn{1}{l|}{0.00000009} &
  0.9229554 \\ \hline
\multicolumn{1}{|l|}{4} &
  \multicolumn{1}{l|}{5,000} &
  \multicolumn{1}{l|}{50} &
  \multicolumn{1}{l|}{0.0004681} &
  \multicolumn{1}{l|}{0.1} &
  \multicolumn{1}{l|}{0.0000008} &
  \multicolumn{1}{l|}{0.00000003} &
  0.7242638 \\ \hline
\multicolumn{1}{|l|}{5} &
  \multicolumn{1}{l|}{3,000} &
  \multicolumn{1}{l|}{100} &
  \multicolumn{1}{l|}{0.0001686} &
  \multicolumn{1}{l|}{0.1} &
  \multicolumn{1}{l|}{0.0000005} &
  \multicolumn{1}{l|}{0.00000007} &
  0.8150428 \\
\hline
\end{longtable}
}

{
\section{Fine-tuning the hyperparameters for the hybrid Estimation of Distribution Algorithm - Continuous Strata}\label{appendixcont}
{
\tiny
\centering
\begin{longtable}{|llllllll|}
\label{HEDAhyper_cont}\\
\caption{Hyperparameters for the HEDA in the case of Continuous Strata}\\
\hline
\multicolumn{8}{|c|}{\textbf{Swiss Municipalities}} \\ \hline
\endfirsthead
\endhead
\multicolumn{1}{|l|}{\textbf{\begin{tabular}[c]{@{}l@{}}Fine-tuning \\ iteration \\ No.\end{tabular}}} &
  \multicolumn{1}{l|}{\textbf{Iterations, I}} &
  \multicolumn{1}{l|}{\textbf{\begin{tabular}[c]{@{}l@{}}SAA \\ Frequency\end{tabular}}} &
  \multicolumn{1}{l|}{\textbf{\begin{tabular}[c]{@{}l@{}}Mutation \\ Chance\end{tabular}}} &
  \multicolumn{1}{l|}{\textbf{\begin{tabular}[c]{@{}l@{}}Elitism \\ Rate\end{tabular}}} &
  \multicolumn{1}{l|}{\textbf{\begin{tabular}[c]{@{}l@{}}Add \\ Strata \\ Factor\end{tabular}}} &
  \multicolumn{1}{l|}{\textbf{\begin{tabular}[c]{@{}l@{}}Temperature, \\ T\end{tabular}}} &
  \textbf{\begin{tabular}[c]{@{}l@{}}Decrement Constant,\\ DC\end{tabular}} \\ \hline
\multicolumn{1}{|l|}{1} &
  \multicolumn{1}{l|}{3,000} &
  \multicolumn{1}{l|}{200} &
  \multicolumn{1}{l|}{0.0000396} &
  \multicolumn{1}{l|}{0.2} &
  \multicolumn{1}{l|}{0.00000082} &
  \multicolumn{1}{l|}{0.0000010} &
  0.6933468 \\ \hline
\multicolumn{1}{|l|}{2} &
  \multicolumn{1}{l|}{1,000} &
  \multicolumn{1}{l|}{200} &
  \multicolumn{1}{l|}{0.0000994} &
  \multicolumn{1}{l|}{0.3} &
  \multicolumn{1}{l|}{0.00000023} &
  \multicolumn{1}{l|}{0.0000043} &
  0.5736595 \\ \hline
\multicolumn{1}{|l|}{3} &
  \multicolumn{1}{l|}{3,000} &
  \multicolumn{1}{l|}{200} &
  \multicolumn{1}{l|}{0.0000292} &
  \multicolumn{1}{l|}{0.3} &
  \multicolumn{1}{l|}{0.00000049} &
  \multicolumn{1}{l|}{0.0000086} &
  0.9229554 \\ \hline
\multicolumn{1}{|l|}{4} &
  \multicolumn{1}{l|}{3,000} &
  \multicolumn{1}{l|}{100} &
  \multicolumn{1}{l|}{0.0000463} &
  \multicolumn{1}{l|}{0.2} &
  \multicolumn{1}{l|}{0.00000076} &
  \multicolumn{1}{l|}{0.0000021} &
  0.7242638 \\ \hline
\multicolumn{1}{|l|}{5} &
  \multicolumn{1}{l|}{1,000} &
  \multicolumn{1}{l|}{200} &
  \multicolumn{1}{l|}{0.0000161} &
  \multicolumn{1}{l|}{0.2} &
  \multicolumn{1}{l|}{0.00000054} &
  \multicolumn{1}{l|}{0.0000066} &
  0.8150428 \\
\hline
\end{longtable}
}}

\end{document}